\documentstyle[11pt,paspconf,epsf]{article}

\begin{document}

\title{Measuring $H_0$ with CLASS B1608+656: The Second Season of
VLA Monitoring}

\author{C. D. Fassnacht}
\affil{NRAO, P. O. Box O, Socorro, NM  87801}

\author{E. Xanthopoulos}
\affil{NRAL Jodrell Bank, University of Manchester, Macclesfield, Cheshire
SK11 9DL, UK}

\author{L. V. E. Koopmans}
\affil{Kapteyn Astronomical Institute, Postbus 800, 9700 Groningen,
The Netherlands}

\author{T. J. Pearson, A. C. S. Readhead}
\affil{Owens Valley Radio Observatory, Caltech, Pasadena, CA 91125}

\author{S. T. Myers}
\affil{NRAO, P. O. Box O, Socorro, NM  87801}

\begin{abstract}
The four-component gravitational lens CLASS B1608+656 has been
monitored with the VLA for two seasons in order to search for time
delays between the components.  These time delays can be combined with
mass models of the lens system to yield a measurement of $H_0$.  The
component light curves show significantly different behavior in the
two observing seasons.  In the first season the light curves have
maximum variations of $\sim$5\%, while in the second season the
components experienced a nearly monotonic $\sim$40\% decrease in flux.
We present the time delays derived from a joint analysis of the light
curves from the two seasons.
\end{abstract}


\section{Introduction}

The four-image gravitational lens CLASS B1608+656 was one of
the first lenses discovered in CLASS (\cite{stm1608}).  Follow-up
radio and optical imaging showed the four-component morphology at all
wavelengths.  In addition, HST multicolor imaging showed the presence
of two lensing galaxies as well as lensed arcs and a faint Einstein
ring (see images in \cite{disklens} and on the CASTLeS web page at
{\tt http://cfa-www.harvard.edu/glensdata/}).  Spectroscopic
observations have given the redshifts of the lensing galaxies and the
lensed background source, which is a post-starburst radio galaxy
(\cite{stm1608}; \cite{cdf1608}).  With the numerous observational
constraints from the HST imaging and evidence of radio variability
from early VLA observations, this system presented an excellent
possibility to be used for a determination of $H_0$.  Thus, a program
of VLA monitoring was started in late 1996.

The results of the first season of VLA monitoring, between 1996
October and 1997 May, are presented in Fassnacht et al.\ (1999).  We
measured the three independent time delays in the system from the
monitoring data, yielding $( \Delta t_{BA}, \Delta t_{BC}, \Delta
t_{BD}) = (31 \pm 7, 36 \pm 7, 76^{+9}_{-10})$ at 95\% confidence.
The time delays were combined with the model presented in Koopmans \&
Fassnacht (1999) to yield $H_0 = 59^{+8}_{-7}$ km~s$^{-1}$~Mpc$^{-1}$
(95\% confidence) with an additional uncertainty of $\pm 15$
km~s$^{-1}$~Mpc$^{-1}$ from the modeling.  The uncertainties on the
measured time delays, and hence on the resulting determination of
$H_0$ are rather large because the background source flux density
varied by only $\sim 5\%$ during the first season of monitoring.  In
principle, stronger variations should allow the alignment of the light
curves more accurately, thus leading to smaller uncertainties in the
measured time delays.  For this reason, a second season of monitoring
was begun.  These proceedings present initial results from the second
season of monitoring.

\section{Observations and Data Reduction}

We observed the B1608+656 system at 8~GHz with the VLA between
February and October 1998.  The system was observed 80 times during
this period, with observations made, on average, every 3.1~d.  With
the exception of using the source 1634+627 as the primary flux
calibrator instead of 3C\,286, the data reduction procedure was the
same as described in Fassnacht et al.\ (1999).  

\section{Results and Discussion}

The final light curves for the four B1608+656 components show large
variations in flux density, on the order of 40\% (Fig.~1a).  
\begin{figure}
\plottwo{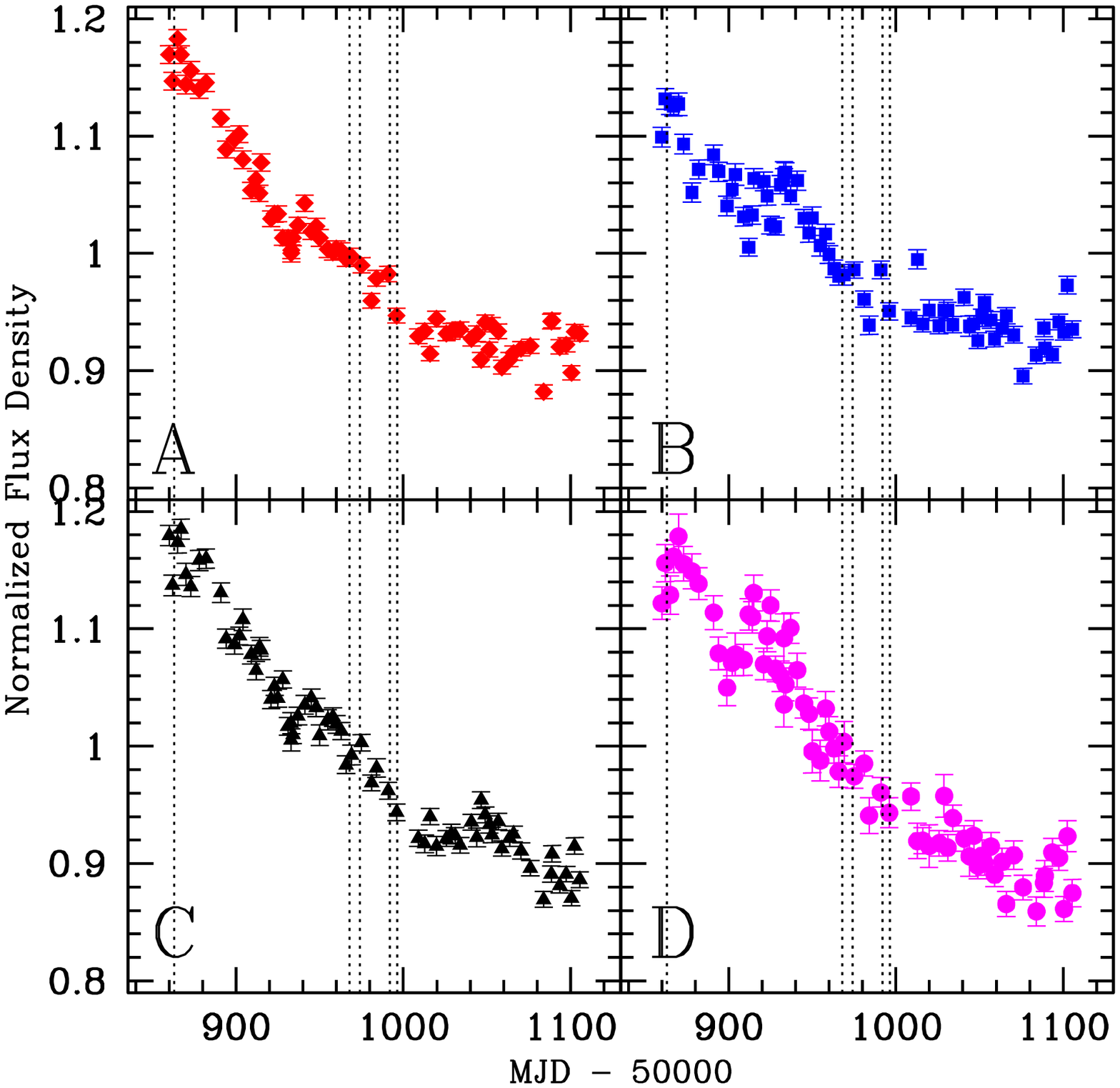}{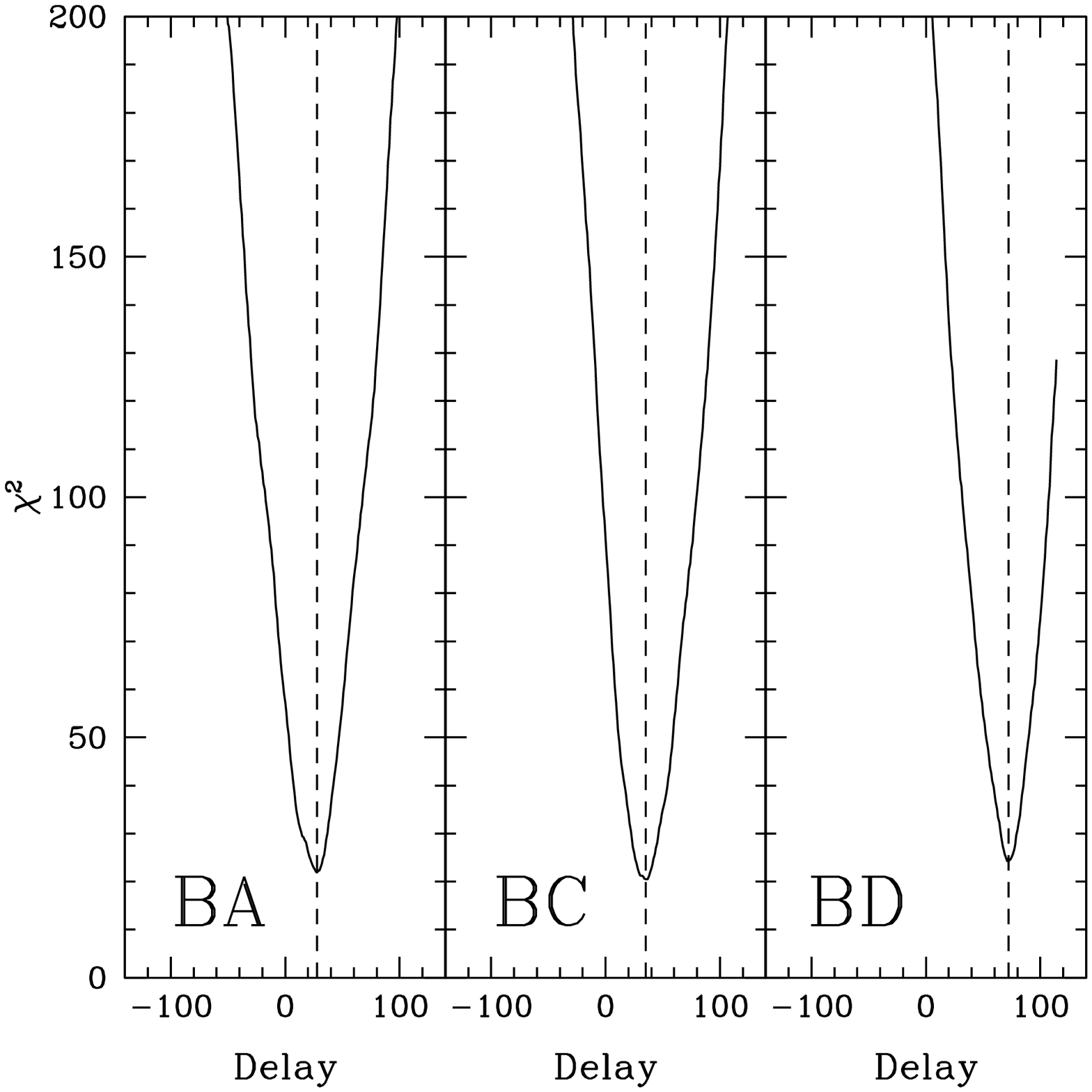}
\caption{
{\bf a) (left)} Component light curves for B1608+656 system from
second season of monitoring.  Each light
curve has been normalized by its mean value (22, 11, 11, and 4~mJy
for components A, B, C, and D, respectively).
{\bf b) (right)} $\chi^2$-minimization curves from the joint analysis
of the first and second seasons' monitoring data.  The minimum reduced 
$\chi^2$ values for this choice of smoothing and interpolation occur
at delays of 28, 35, and 72~d.}
\end{figure}
These variations are quite a contrast to the variations of only
$\sim$5\% seen in the light curves from the first season of
monitoring.  However, the nature of the variations, a nearly monotonic
decline, introduces a degeneracy in magnification-delay space, making
it difficult to align the curves accurately.  In order to break this
degeneracy, we have conducted a joint analysis of the data from the
first two seasons.  This approach is useful because the nearly
constant light curves from the first season limit the allowed range in
magnification while the steeply declining second season curves limit
the allowed range of delays.  An initial $\chi^2$ minimization
analysis (see Fassnacht et al.\ 1999 for a description of the details)
produces much clearer minima than those produced from the first-season
data alone (Fig.~1b).  We also conducted Monte Carlo simulations to
estimate the uncertainties on the delays.  As hoped, the uncertainties
decreased by 30 -- 50\%.  The delays produced by this initial analysis
are thus $(\Delta t_{BA}, \Delta t_{BC}, \Delta t_{BD}) = (26 \pm 5,
34 \pm 5, 73 \pm 5)$ at 95\% confidence.  The delays derived from the
joint analysis are consistent within the errors with those produced
from the first season.  However, there is a systematic shift to lower
delays, which is especially noticeable for the shortest delay.  This
change may be due to small errors in matching the absolute flux
densities of the curves between the two seasons, an effect that is
being investigated further.  Composite curves created by shifting the
component light curves by these delays are presented in Fig.~2.
\begin{figure}
\plottwo{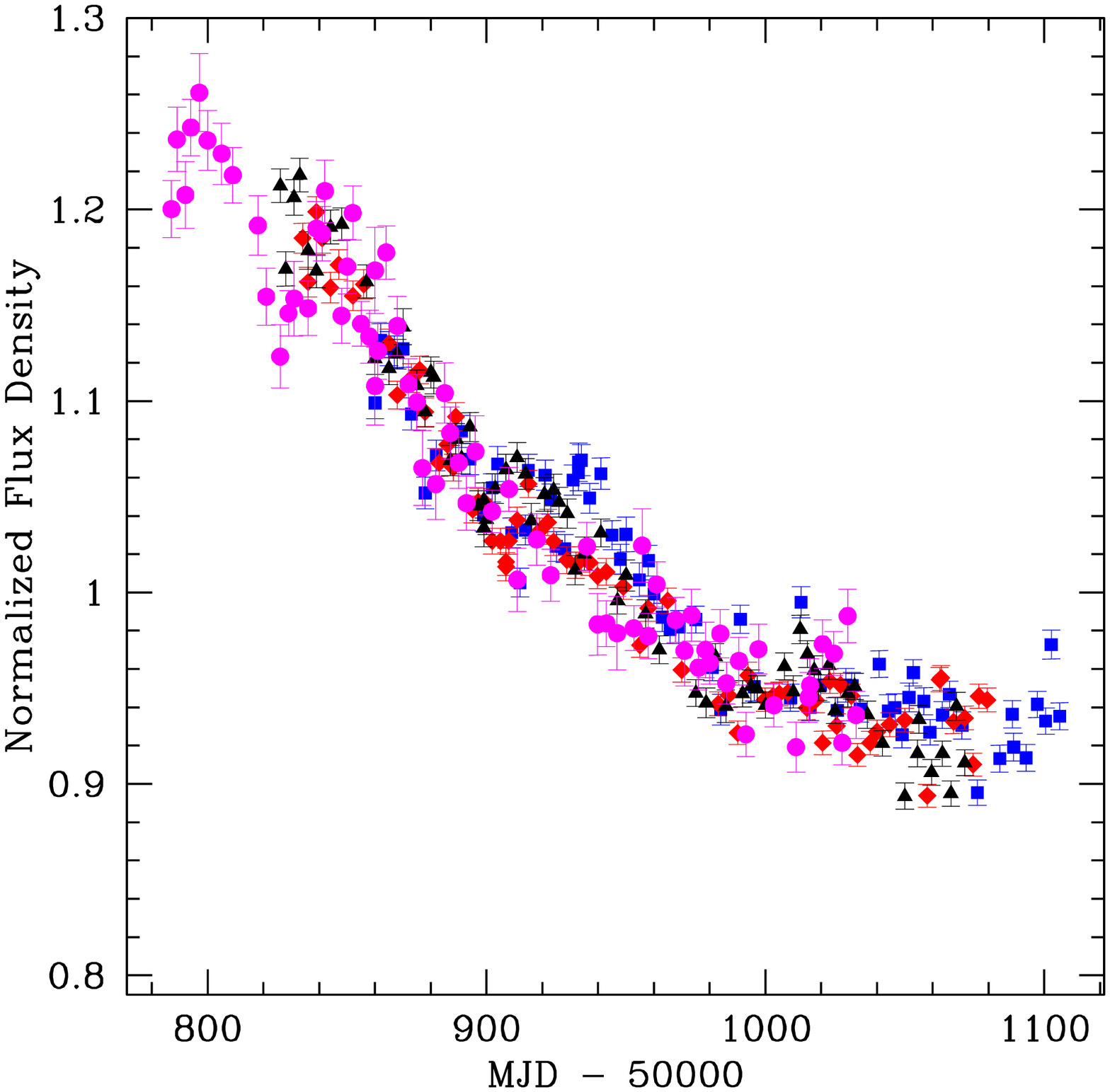}{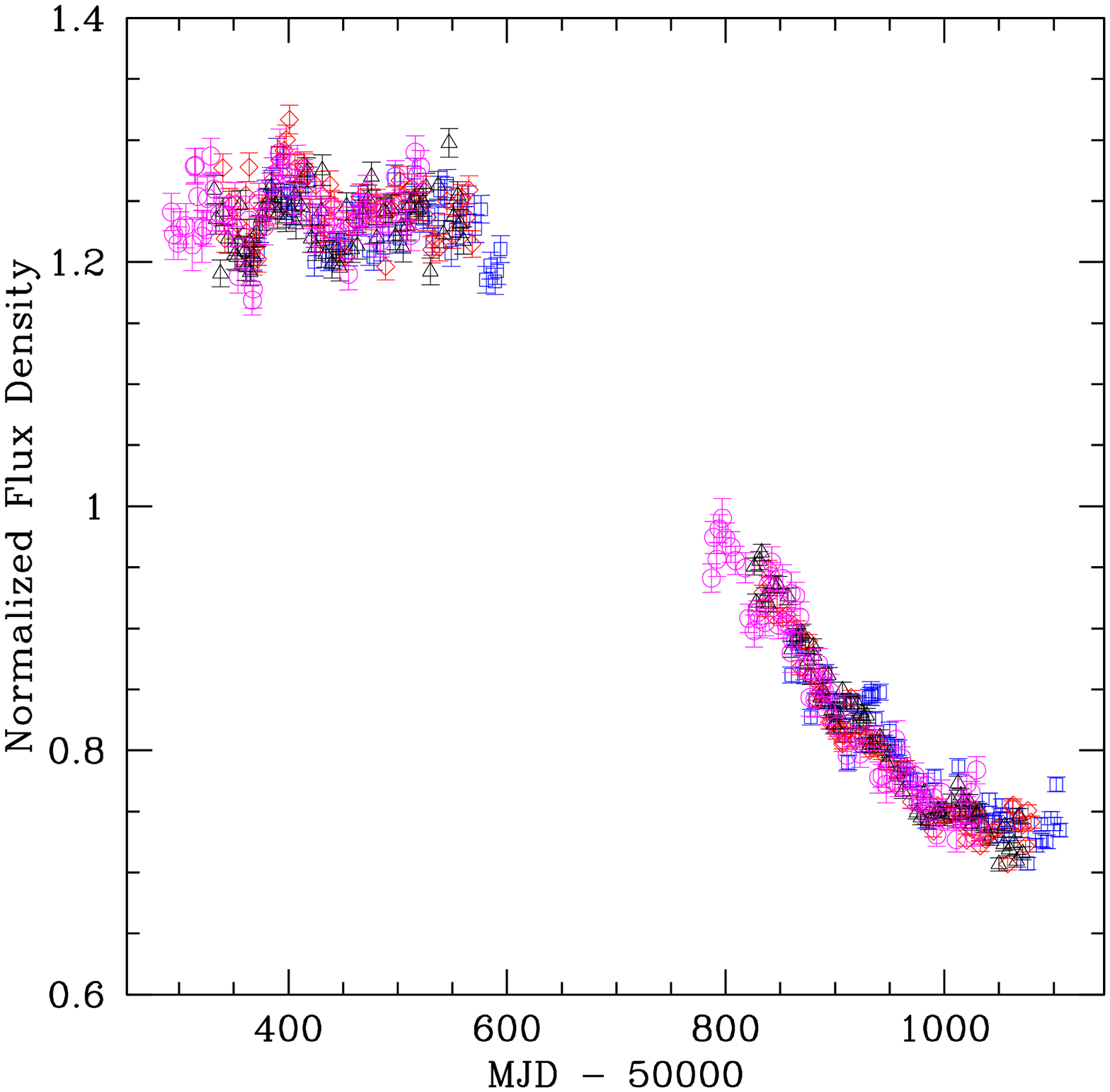}
\caption{Composite curves created by shifting the component light
curves by the measured time delays and overlaying them.  Symbols
as in Fig.~1a. {\bf a) (left)} Season 2 only. {\bf b) (right)}
Season 1 and season 2.}
\end{figure}

\section{Future Prospects}

The B1608+656 system has produced a measurement of $H_0$ through a
combination of the measured time delays with a mass model for the
lensing galaxies.  However, the uncertainties in both the lens model
and the measured delays are still larger than desired.  Future work on
this system could reduce the uncertainties in both of these areas.  On
the modeling side, Blandford and Surpi are developing a new lens model
using the full information in the HST images of the system.  The time
delay uncertainties also can be reduced if the source varies in a
strong, but non-monotonic, manner.  It should be noted that just such
a variation occurred during the gap between the first two seasons when
the source began its change from its previous, nearly constant, flux
density.  This change in behavior was missed because the VLA was in
its compact configurations and could not resolve the individual source
components.  We are currently conducting a third season of monitoring
with the VLA to search for more variability in the background source.

\acknowledgments

The National Radio Astronomy Observatory is a facility of the National
Science Foundation operated under cooperative agreement by Associated
Universities Inc.  This work is supported in part by the NSF under
grant \#AST 9420018 and by the European Commission, TMR Program,
Research Network Contract ERBFMRXCT96-0034 ``CERES.''

\end{document}